\documentclass{article}

\title{ \bf Quark-Lepton Masses and the Neutrino Puzzle in 
the AGUT Model}
\author{Colin D Froggatt, \\[0.2cm]
\it Department of Physics and Astronomy, Glasgow University,\\
\it Glasgow G12 8QQ, Scotland, UK}
\begin{document}
\begin{flushright}
GUTPA/01/12/02\\
\end{flushright}
\vskip .1in
\begin{center}

{\LARGE \bf Quark-Lepton Masses and the Neutrino \\ 
\vspace{6pt}
Puzzle in the AGUT Model}

\vspace{20pt}

{\bf \large Colin D. Froggatt}

\vspace{6pt}

{ \em Department of Physics and Astronomy\\
 Glasgow University, Glasgow G12 8QQ,
Scotland\\}
\end{center}
\vspace{6pt}
\section*{ }
\begin{center}
{\large\bf Abstract}
\end{center}

We first discuss an approach to the fermion mass problem, 
according to which the whole of flavour
mixing for quarks is determined by the mechanism responsible for
generating the physical masses of the {\it up} and {\it down} quarks:
the Lightest Flavour Mass Generation model. Then we consider 
fermion masses in the Anti-grand Unification Theory and, 
in particular, the neutrino mass and mixing problem. 

\vspace{200pt}

To be published in the 
Proceedings of the International Workshop on 
{\it What comes beyond the Standard Model}, Bled, Slovenia, 
July 2001.

\thispagestyle{empty}
\newpage

\date{}
\maketitle

\section{Introduction}
I reviewed the general problem of the quark-lepton 
mass spectrum at the first Bled workshop on ``What 
comes beyond the Standard Model'' \cite{bled1}. 
So, in this talk, I will mainly concentrate on two 
topics: the Lightest Flavour Mass Generation model 
and the Neutrino Mass and Mixing problem in the 
Anti-Grand Unification Theory (AGUT). 


\section{Lightest Flavour Mass Generation Model}

A commonly accepted framework for discussing the flavour problem is based on
the picture that, in the absence of flavour mixing, only the particles
belonging to the third generation $t$, $b$ and $\tau $ have non-zero masses.
All other masses and the mixing angles then appear as a result of
the tree-level mixings of families, related to some underlying family
symmetry breaking. 
Recently, a new mechanism of flavour mixing, which we call Lightest Family
Mass Generation (LFMG), was proposed \cite{lfm}. 
According to LFMG the whole of flavour
mixing for quarks is basically determined by the mechanism responsible for
generating the physical masses of the {\it up} and {\it down} quarks, 
$m_{u}$ and $m_{d}$ respectively. So, in the chiral symmetry limit, when 
$m_{u}$ and $m_{d}$ vanish, all the quark mixing angles vanish. 
Therefore, the masses (more precisely
any of the diagonal elements of the quark and
charged lepton mass matrices)
of the second and third families are practically the same in
the gauge (unrotated) and physical bases.
The proposed flavour mixing mechanism, driven solely by the generation of
the lightest family mass, could actually be realized in two generic ways.

The first basic alternative (I) is when the lightest family mass ($m_{u}$ or 
$m_{d} $) appears as a result of the complex flavour mixing of all three
families. It ``runs along the main diagonal'' of the corresponding $3\times
3 $ mass matrix $M$, from the basic dominant element $M_{33}$ to the element 
$M_{22}$ (via a rotation in the 2-3 sub-block of $M$) and then to the
primordially texture zero element $M_{11}$ (via a rotation in the 1-2
sub-block). The direct flavour mixing of the first and third quark and 
lepton families is supposed to be absent or negligibly small in $M$.

The second alternative (II), on the contrary, presupposes direct flavour
mixing of just the first and third families. There is no involvement of the
second family in the mixing. In this case, the lightest mass appears in the
primordially texture zero $M_{11}$ element ``walking round the corner'' (via
a rotation in the 1-3 sub-block of the mass matrix $M$). Certainly, this
second version of the LFMG mechanism cannot be used for both the up and the
down quark families simultaneously, since mixing with the second family
members is a basic part of the CKM quark mixing 
phenomenology (Cabibbo mixing, non-zero $V_{cb}$ 
element, CP violation). However, the alternative II could work for
the up quark family provided that the down quarks follow the alternative I.

Here we will just consider the latter scenario.

\subsection{Quark Sector}
We propose that the mass matrix for the down quarks ($D$ = $d$, $s$, $b$) 
is Hermitian with three texture zeros of the following 
alternative I form:

\begin{equation}
M_{D}=\pmatrix{ 0 & a_D & 0 \cr a_D^{\ast} & A_D & b_D \cr 0 & 
b_D^{\ast} & B_D \cr}  \label{LFM1}
\end{equation}
It is, of course, necessary to assume some hierarchy between the elements,
which we take to be: $B_{D}\gg A_{D}\sim \left| b_{D}\right| \gg \left|
a_{D}\right| $. The zero in the $\left( M_{D}\right)
_{11}$ element corresponds to the commonly accepted conjecture
that the lightest family masses appear as a direct result of flavour
mixings. The zero in $\left( M_{D}\right) _{13}$ means that only minimal
``nearest neighbour'' interactions occur, giving a tridiagonal matrix
structure.

Now our main hypothesis, that the second and third family diagonal mass
matrix elements are practically the same in the gauge and physical
quark-lepton bases, means that : 
\begin{equation}
B_{D}=m_{b}+\delta_{D} \qquad A_{D}= m_{s} + \delta_{D}^{\prime }  
\label{BA}
\end{equation}
The components $\delta_{D} $ and $\delta_{D}^{\prime }$ are supposed to be
much less than the masses of the particles in the next lightest family,
meaning: 
\begin{equation}
|\delta_D |\ll m_{s} \qquad |\delta _{D}^{\prime }|\ll m_{d}  
\label{deldelp}
\end{equation}
Since the trace and determinant of the Hermitian matrix $M_{D}$ gives the
sum and product of its eigenvalues, it follows that 
\begin{equation}
\delta _{D}\simeq - m_{d}  \label{del}
\end{equation}
while $\delta _{D}^{\prime }$ is vanishingly small and can be neglected
in further considerations.

It may easily be shown that our hypothesis and related equations (\ref{BA} - 
\ref{del}) are entirely equivalent to the condition that the diagonal
elements ($A_{D}$, $B_{D}$) of $M_{D}$ are proportional
to the modulus square of the off-diagonal elements ($a_{D}$, $b_{D}$): 
\begin{equation}
\frac{A_{D}}{B_{D}}=\left| \frac{a_{D}}{b_{D}}\right| ^{2}
\label{ABab}
\end{equation}

Using the conservation of the trace, determinant and sum of principal minors
of the Hermitian matrices $M_{D}$ under unitary transformations, we are led
to a complete determination of the moduli of all their elements, which
can be expressed to high accuracy as follows: 
\begin{equation}
\left| M_{D} \right| = \pmatrix{ 0 & \sqrt{m_d m_s} & 0 \cr 
\sqrt{m_d m_s} & m_s & \sqrt{m_d m_b} \cr 0 & 
\sqrt{m_d m_b} & m_b - m_d \cr}  \label{LFM1A}
\end{equation}

Now the Hermitian mass matrix for the up quarks is taken to be 
of the following alternative II form: 
\begin{equation}
M_{U}=\pmatrix{ 0 & 0 & c_U \cr 0 & A_U & 0 \cr c_U^{\ast} & 0 & B_ U \cr}
\label{LFM2}
\end{equation}
The moduli of all the elements of $M_{U}$ can also be readily 
determined in terms of the physical masses as follows: 
\begin{equation}
\left| M_{U} \right| = \pmatrix{ 0 & 0 & \sqrt{m_u m_t} \cr 
0 & m_c & 0 \cr \sqrt{m_u m_t} & 0 & m_t - m_u \cr}
\label{LFM2A}
\end{equation}

The CKM quark mixing matrix elements can now be readily calculated 
by diagonalising the mass matrices $M_D$ and $M_U$. They are 
given by the following simple and compact formulae in terms of 
quark mass ratios:  
\begin{eqnarray}
\left| V_{us}\right| = \sqrt{\frac{m_{d}}{m_{s}}} = 0.222 \pm 0.004 
\qquad \left| V_{us}\right|_{exp} = 0.221 \pm 0.003 \\
\left|V_{cb}\right| = \sqrt{\frac{m_{d}}{m_{b}}} = 0.038 \pm 0.004 
\qquad \left|V_{cb}\right|_{exp} = 0.039 \pm 0.003 \\ 
\left|V_{ub}\right| = \sqrt{\frac{m_{u}}{m_{t}}} = 0.0036 \pm 0.0006
\qquad \left|V_{ub}\right|_{exp} = 0.0036 \pm 0.0006   \label{angles}
\end{eqnarray}
As can be seen, they are in impressive agreement with the experimental 
values.

\subsection{Lepton Sector}

The MNS lepton mixing matrix is defined analogously to the CKM 
quark mixing matrix: 
\begin{equation}
U=U_{\nu }U{_{E}}^{\dagger }  \label{U}
\end{equation}
Here $U_E$ and $U_{\nu}$ are the unitary matrices which diagonalise 
the charged lepton mass matrix $M_E$ and the effective neutrino mass 
matrix $M_{\nu}$ respectively. Assuming the charged lepton masses 
follow alternative I, like the down quarks, the LFMG model 
predicts the charged lepton mixing angles in the matrix $U_{E}$ 
to be: 
\begin{equation}
\sin \theta _{e\mu }=\sqrt{\frac{m_{e}}{m_{\mu }}}\qquad \sin \theta _{\mu
\tau }=\sqrt{\frac{m_{e}}{m_{\tau }}}\qquad \sin \theta _{e\tau }\simeq 0
\label{charged}
\end{equation}
These small charged lepton mixing angles will not
markedly effect atmospheric neutrino oscillations, which
appear to require maximal mixing $\sin ^{2}2\theta _{atm}\simeq
1 $. Similarly, in the case of the large mixing angle (LMA) MSW solution  
of the solar neutrino problem, they are essentially negligible. 
It follows then that the large neutrino mixings should mainly come 
from the $U_{\nu }$
matrix associated with the neutrino mass matrix. 

According to the ``see-saw''
mechanism, the effective mass-matrix $M_{\nu}$ for physical neutrinos has the
form 
\begin{equation}
M_{\nu }=-M_{N}^{T}M_{NN}^{-1}M_{N}  \label{nu}
\end{equation}
where $M_{N}$ is their Dirac mass matrix, while $M_{NN}$ is the Majorana
mass matrix of their right-handed components. 
Matsuda et al \cite{refs} have extended the alternative I LFMG texture 
to the Dirac $M_N$ and Majorana $M_{NN}$ matrices. 

The eigenvalues of the neutrino Dirac mass matrix $M_{N}$ are
taken to have a hierarchy similar to that for the charged leptons (and down
quarks)

\begin{equation}
M_{N3}:M_{N2}:M_{N1}\simeq 1:y^{2}:y^{4},\quad y\approx 0.1  \label{h1}
\end{equation}
and the eigenvalues of the Majorana mass matrix $M_{NN}$ are taken to have a
stronger hierarchy

\begin{equation}
M_{NN3}:M_{NN2}:M_{NN1}\simeq 1:y^{4}:y^{6}  \label{h2}
\end{equation}
One then readily determines the general LFMG matrices $M_{N}$ and $M_{NN}$
to be of the type

\begin{equation}
M_{N}\simeq M_{N3}\left( 
\begin{array}{lll}
0 & \alpha y^{3} & 0 \\ 
\alpha y^{3} & y^{2} & \alpha y^{2} \\ 
0 & \alpha y^{2} & 1
\end{array}
\right)  \label{m0}
\end{equation}
and

\begin{equation}
M_{NN}\simeq M_{NN3}\left( 
\begin{array}{lll}
0 & \beta y^{5} & 0 \\ 
\beta y^{5} & y^{4} & \beta y^{3} \\ 
0 & \beta y^{3} & 1
\end{array}
\right).  \label{m00}
\end{equation}
We further take an extra condition of
the type

\begin{equation}
\left| \Delta -1\right| \le y^{2}\qquad (\Delta \equiv
\alpha ,\beta )  \label{natur}
\end{equation}
for both the order-one parameters $\alpha $ and $\beta $ contained 
in the matrices $M_{N}$ and $M_{NN}$, 
according to which they are supposed to be equal to unity with a few percent
accuracy. Substitution in the seesaw formula (\ref{nu}) generates an
effective physical neutrino mass matrix $M_{\nu }$ of the form:

\begin{equation}
M_{\nu }\simeq -\frac{M_{N3}^{2}}{M_{NN3}}\left( 
\begin{array}{lll}
0 & y & 0 \\ 
y & 1+(y-y^{2})^{2} & 1-(y-y^{2}) \\ 
0 & 1-(y-y^{2}) & 1
\end{array}
\right)  \label{matr1}
\end{equation}
The physical neutrino masses are then given by:
\begin{eqnarray}
m_{\nu 1} &\simeq &(\frac{1}{2}-\frac{\sqrt{3}}{2})\frac{M_{N3}^{2}}{M_{NN3}}%
\cdot y, \nonumber \\
m_{\nu 2} &\simeq &(\frac{1}{2}+\frac{\sqrt{3}}{2})\frac{M_{N3}^{2}}{M_{NN3}}%
\cdot y,\quad m_{\nu 3}\simeq (2-y)\frac{M_{N3}^{2}}{M_{NN3}}  
\end{eqnarray}
The predicted values of the neutrino oscillation parameters are: 
\begin{equation}
\sin ^{2}2\theta _{atm}\simeq 1,\quad \sin ^{2}2\theta _{sun}\simeq \frac{2}{%
3},\quad U_{e3}\simeq \frac{1}{2\sqrt{2}}y,\quad \frac{\Delta m_{sun}^{2}}{%
\Delta m_{atm}^{2}}\simeq \frac{\sqrt{3}}{4}y^{2}  \label{pred1}
\end{equation}
in agreement with atmospheric and LMA-MSW solar neutrino oscillation 
data.

The proportionality condition (\ref{ABab}), which leads to the LFMG 
texture, is not so easy to generate 
from an underlying symmetry beyond the Standard Model. 
However Jon Chkareuli, Holger Nielsen and myself have recently shown 
\cite{SU3} that it is possible to give a natural realisation of the 
LFMG texture in a local chiral $SU(3)$ family symmetry model. 

\section{Fermion Masses in the AGUT Model}

The AGUT model is based on a non-simple extension of the Standard 
Model (SM) with three copies of the SM gauge group---one for each 
family---and, in the absence of right-handed neutrinos, 
one extra abelian factor: $G = SMG^3 \times U(1)_f$, where 
$SMG \equiv SU(3) \times SU(2) \times U(1)$. This AGUT gauge 
group is broken down by four Higgs fields $S$, $W$, $T$ and 
$\xi$ to the usual SM gauge group, identified as the 
diagonal subgroup of $SMG^3$. The Higgs field $S$ has a 
vacuum expectation value (VEV) taken to be unity in fundamental 
(Planck) mass units, while $W$, $T$ and $\xi$ have VEVs an 
order of magnitude smaller. So the pure SM is essentially 
valid, without supersymmetry, up to energies close to the 
Planck scale. The AGUT gauge group $SMG^3 \times U(1)_f$  
only becomes effective near the Planck scale, where the 
$i$'th proto-family couples to just the $i$'th $SMG$ factor 
and $U(1)_f$. The $U(1)_f$ charges assigned to the quarks 
and leptons are determined, by anomaly cancellation constraints, 
to be zero for the first family and all left-handed fermions, 
and for the remaining right-handed states to be as follows:
\begin{equation}
Q_f(\tau_r) = Q_f(b_R) = Q_f(c_R) =1 \qquad 
Q_f(\mu_r) = Q_f(s_R) = Q_f(t_R) = -1
\end{equation}   
I refer to the review of the AGUT model by Holger and myself 
at the first Bled workshop \cite{bled2} for more details.

The quarks and leptons are mass protected by the approximately 
conserved AGUT chiral gauge charges \cite{fn}. The quantum numbers 
of the Weinberg-Salam Higgs field $\phi_{WS}$ are chosen so 
that the $t$ quark mass is not suppressed, whereas the $b$ 
quark and $\tau$ lepton are suppressed. This is done by 
taking the four abelian charges, expressed as a charge 
vector $\vec{Q} = (y_1/2, y_2/2, y_3/2, Q_f)$, for $\phi_{WS}$ 
to be given by:
\begin{equation}
\vec{Q}_{\phi_{WS}} = \vec{Q}_{c_R} - \vec{Q}_{t_L} 
=(0,2/3,0,1) - (0,0,1/6,0) = (0,2/3,-1/6,1) 
\end{equation}
We assume that, like the quark and lepton fields, the Higgs 
fields belong to singlet or fundamental representations of 
all the non-abelian groups. Then, by imposing the usual SM 
charge quantisation rule for each of the $SMG$ factors, the 
non-abelian representations are determined from the weak 
hypercharge quantum numbers $y_i$. The abelian quantum 
numbers of the other Higgs fields are chosen as follows:
\begin{eqnarray}
\vec{Q}_W = (0,-1/2,1/2,-4/3) \qquad \vec{Q}_T = 
(0,-1/6,1/6,-2/3) \\
\vec{Q}_{\xi} = (1/6,-1/6,0,0) \qquad \vec{Q}_S = 
(1/6,-1/6,0,-1)
\end{eqnarray}   
Since we have $<S>=1$ in Planck units, the Higgs field $S$ 
does not suppress the fermion masses and the quantum numbers 
of the other Higgs fields $W$, $T$, $\xi$ and $\phi_{WS}$ 
given above are only determined modulo those of $S$.

The effective SM Yukawa coupling matrices in this AGUT model can 
now be calculated in terms of the VEVs of the fields $W$, $T$ 
and $\xi$ in Planck units---up to ``random`` complex order unity 
factors multiplying all the matrix elements---for the quarks:
\begin{equation}
Y_U \sim \left ( \begin{array}{ccc}
        W T^2 \xi^2 & W T^2 \xi & W^2 T \xi \\
        W T^2 \xi^3 & W T^2 & W^2 T\\
        \xi^3  & 1 & W T
        \end{array} \right ) \label{H_U} \quad
Y_D  \sim  \left ( \begin{array}{ccc}
        W (T^2 \xi^2 & W T^2 \xi & T^3 \xi \\
        W T^2 \xi & W T^2 & T^3 \\
        W^2 T^4 \xi & W^2 T^4 & W T 
	\end{array} \right ) \label{H_D} 
\end{equation}
and the charged leptons:
\begin{equation}
Y_E  \sim  \left ( \begin{array}{ccc}
	W T^2  \xi^2 & W T^2 \xi^3 & W T^4 \xi \\
	W T^2 \xi^5 & W T^2 & W T^4 \xi^2 \\
        W T^5 \xi^3 & W^2 T^4 & W T
        \end{array} \right ) \label{H_E}
\end{equation}
A good order of magnitude fit is then obtained \cite{bled2} 
to the charged fermion masses with the following values for 
the Higgs field VEVs in Planck units:
\begin{equation}
W = 0.179, \qquad T = 0.071 \qquad \xi = 0.099.
\end{equation}

We now consider the neutrino mass matrix in the AGUT model.

\section{Neutrino Mass and Mixing Problem} 

Without introducing new physics below the AGUT scale, the 
effective light neutrino mass matrix $M_{\nu}$ is generated 
by tree level diagrams involving the exchange of two 
Weinberg-Salam Higgs tadpoles and the appropriate 
combination of $W$, $T$, $\xi$ and $S$ Higgs tadpoles. 
In this way we obtain:
\begin{equation}
\label{eq:Mnuminagut}
M_{\nu} \simeq \frac{\langle {\phi}_{WS} \rangle^2}{M_{Pl}} 
\left ( \begin{array}{ccc}
        W^2 {\xi}^4 T^4  & W^2 {\xi} T^4  & W^2 {\xi}^3 T \\
        W^2 \xi T^4  & W T^5 & W^2 T \\
        W^2 {\xi}^3 T  & W^2 T & W^2 T^2 {\xi}^2 
\end{array} \right ),
\end{equation}
The off-diagonal element $(M_{\nu})_{23} = (M_{\nu})_{32}$ dominates 
the matrix, giving large $\nu_{\tau}-\nu_{\mu}$ mixing with the 
following two neutrino masses and mixing angle:
\begin{equation}
m_2 \sim m_3 \sim \frac{\langle {\phi}_{WS} \rangle^2}{M_{Planck}}W^2T 
\qquad \sin^2 2\theta_{\mu \tau} \simeq 1
\end{equation} 
Although the large mixing angle $\sin^2 2\theta_{\mu \tau}$ is 
suitable for atmospheric neutrino oscillations, there are two 
problems associated with the neutrino masses. Firstly the ratio of 
neutrino mass squared differences $\Delta m_{23}^2/\Delta m_{12}^2 
\sim 2 T \xi^2 \sim 1.4 \times 10^{-3}$, whereas the small mixing 
angle (SMA) MSW solution to the solar neutrino problem requires 
$\Delta m_{23}^2/\Delta m_{12}^2 \sim 10^{-2}$. Secondly the 
predicted overall absolute mass scale for the neutrinos 
$\langle {\phi}_{WS} \rangle^2/M_{Planck} \sim 3 \times 10^{-6}$ eV 
is far too small. 

We conclude it is necessary to introduce a new mass scale into the 
AGUT model. Two ways have been suggested of obtaining realistic 
neutrino masses and mixings in the AGUT model:
\begin{enumerate}
\item By extending the AGUT Higgs spectrum to include a weak 
isotriplet Higgs field $\Delta$ with SM weak hypercharge 
$y/2 =-1$ and a VEV $\langle \Delta^0 \rangle \sim 1$ eV; 
also a new Higgs field $\psi$ giving large $\mu - \tau$ mixing 
in the charged lepton Yukawa coupling matrix $Y_E$ is required.
\item By including right-handed neutrinos and extending the AGUT 
gauge group to $G_{extended} = (SMG \times U(1)_{B-L})^3$; also 
two new Higgs fields $\phi_{B-L}$ and $\chi$ are introduced to 
provide a see-saw mass scale and structure to the Majorana 
right-handed neutrino mass matrix. 
\end{enumerate}

Yasutaka Takanishi reported on the second approach 
\cite{yasutaka} at the workshop; so I will report on the 
first approach \cite{mark} here. 
We must therefore consider the introduction of a new Higgs field 
$\psi$, which can yield large mixing from the charged lepton mass 
matrix without adversely affecting the quark mass matrices.
With the following choice of charges for the $\psi$ field
\begin{equation}
\vec{Q}_{\psi}  =  3\vec{Q}_{\xi} + \vec{Q}_W + 4\vec{Q}_T 
 =  \left ( \frac{1}{2}, -\frac{5}{3}, \frac{7}{6}, -4
\right),
\end{equation}
we obtain new expresssions for the quark Yukawa matrices:
\begin{equation}
\label{eq:psimat}
Y_U  =  \left(\begin{array}{ccc} WT^2\xi^2 & WT^2\xi & W^2T\xi \\
                                   WT^2\xi^3 & WT^2    & W^2T    \\
                                   \xi^3     & 1       & WT \end{array}
          \right) \quad
Y_D  =  \left(\begin{array}{ccc} WT^2\xi^2 & WT^2\xi & T^3\xi  \\
                                   WT^2\xi   & WT^2    & T^3     \\
                                   W^2T\psi & W^2T\xi\psi  & WT \end{array}
          \right) \label{eq:H_DLSND}
\end{equation}
and the charged lepton Yukawa matrix:
\begin{equation}
Y_E  =  \left(\begin{array}{ccc} WT^2\xi^2 & W^2T^2\psi & \xi^4\psi \\
                                   W^4T\xi\psi^2 & WT^2    & \xi\psi \\
                                   W^3\xi^2\psi & W^2T\xi\psi  & WT 
\end{array} \right).
\end{equation}
As we can see from the charged lepton matrix we will indeed have
large mixing if $\left\langle \psi \right\rangle = O(0.1)$, so that
$(Y_E)_{23} \sim (Y_E)_{33}$. In the following
discussion we shall take $\psi$ to have a vacuum expectation
value of $\left\langle \psi \right\rangle = 0.1$ for definiteness. 
The effect of the field $\psi$
on the charged fermion masses is then small, since the elements involving 
$\psi$
do not make any significant contribution to the determinant, or the sum
of the minors, of the mass matrices, or the trace of the squares
$Y Y^{\dagger}$
of the Yukawa matrices. The mixings of the quarks is essentially unaffected
by the terms involving $\psi$, and the only 
significant effect is on the mixing matrix $U_E$, which is now given by:
\begin{equation}
\label{eq:UEX}
U_E  \sim  \left( \begin{array}{ccc} 
1 & \frac{\xi \psi^2X}{T} &  \frac{\xi^3}{X} \\
-W\psi & \frac{WT}{\xi\psi X} & \frac{1}{X} \\
\frac{\xi \psi^2}{T} & -\frac{1}{X} & \frac{WT}{\xi\psi X}
\end{array}\right)
 \sim  \left( \begin{array}{ccc}
1 & 0.021 & 6.4 \times 10^{-4} \\
- 0.016 & 0.75 & 0.66 \\
0.014 & - 0.66 & 0.75 \end{array}\right)
\end{equation}
where
$X = \sqrt{1 + W^2T^2/\xi^2 \psi^2} \sim 1.51$
This gives the large mixing required, $\sin^2 2\theta_{atm} 
\sim 1$, for the atmospheric neutrino oscillations.

We can further obtain a solution with vacuum oscillations
for the solar neutrinos by choosing appropriate charges for 
the isotriplet Higgs field $\Delta$.
We require a large off-diagonal $(1, 2)$ element for the neutrino mass
matrix and hence we choose the charges on $\Delta$ to be
\begin{equation}
 \vec{Q}_{\Delta} = ( -\frac{1}{2}, -1,\frac{1}{2}, \frac{5}{3})
\end{equation}
We then obtain the neutrino mass matrix,
\begin{equation}
M_{\nu} \sim \langle\Delta^0\rangle
\left ( \begin{array}{ccc} W \xi^6 &  W  \xi^3 & T \xi^2 \psi\\
W \xi^3 & W & T \xi \psi\\
 T \xi^2 \psi & T \xi \psi &  T^2 \xi \psi \end{array} \right).
\end{equation}
This has the eigenvalues,
\begin{equation}
m_1  \sim   \langle \Delta^0 \rangle \left(-T \xi^2 \psi + 
\frac{T^2 \xi \psi}{2} \right),\quad
m_2  \sim  \langle \Delta^0 \rangle W, \quad
m_3  \sim  \langle \Delta^0 \rangle \left( T \xi^2 \psi + 
\frac{T^2 \xi \psi}{2}
\right)
\end{equation}
where the splitting between $m_1$ and $m_2$ comes from the
mass matrix element $(M_{\nu})_{33}$.  
The neutrino mixing matrix is then given by,
\begin{equation}
U_{\nu} \sim \left ( \begin{array}{ccc}
\frac{1}{\sqrt{2}}(1 + \frac{T}{4\xi}) & \xi^3 &
\frac{1}{\sqrt{2}}(1 - \frac{T}{4\xi}) \\
\frac{T\xi\psi}{\sqrt{2}W}(1 - \frac{T}{4\xi}) & 1 &
-\frac{T\xi\psi}{\sqrt{2}W}(1 + \frac{T}{4\xi}) \\
-\frac{1}{\sqrt{2}}(1 - \frac{T}{4\xi}) & \frac{T\xi\psi}{W} &
\frac{1}{\sqrt{2}}(1 + \frac{T}{4\xi})
\end{array} \right).
\end{equation}
Hence, using  $U_E$ from eqn.~\ref{eq:UEX}, we have the 
lepton mixing matrix $U = U_E^{\dagger} U_{\nu}$:  
\begin{equation}
U \sim \left ( \begin{array}{ccc}
\frac{1}{\sqrt{2}}(1 + \frac{T}{4\xi}) & -W\xi &
\frac{1}{\sqrt{2}}(1 - \frac{T}{4\xi})\\
\frac{1}{\sqrt{2}X}(1 - \frac{T}{4\xi}) & \frac{WT}{\xi \psi X} &
-\frac{1}{\sqrt{2}}(1 + \frac{T}{4\xi})\\
-\frac{WT(1 - \frac{T}{4\xi})}{\sqrt{2}\xi\psi X} & \frac{1}{X} &
\frac{WT(1 + \frac{T}{4\xi})}{\sqrt{2}\xi\psi X}
\end{array} \right) 
\sim  \left ( \begin{array}{ccc}
0.83 & -0.016 & 0.58\\
0.38 & 0.75 & -0.55\\
-0.43 & 0.66 & 0.62
\end{array} \right)
\end{equation}
which, as we can see, has large electron neutrino mixing, as we require
for a vacuum oscillation solution to the solar neutrino problem.  
 
We also have the mass hierarchy,
\begin{equation}
\frac{\Delta m^2_{13}}{\Delta m^2_{23}} \sim 2 \frac{T^3 \xi^3 \psi^2}{W^2}
\sim 3 \times 10^{-7}.
\end{equation}
Hence, if we then take $\langle \Delta^0 \rangle \sim 0.2$ eV, 
so that we have an
overall mass scale suitable for the atmospheric neutrino problem, then we
will also have,
\begin{equation}
\Delta m^2_{13} \sim 2 \langle\Delta^0\rangle^2 T^3 \xi^3 \psi^2 \sim
3 \times 10^{-10} \mbox{eV}^2.
\end{equation}
With such a hierarchy of $\Delta m^2$s we effectively
have two-neutrino oscillations for the solar neutrinos, with
the mixing angle given by,
\begin{equation}
\sin^2 2\theta_{sun} = 4 U^2_{e1} U^2_{e3} \sim 0.9.
\end{equation}
So, we have the `just-so' vacuum oscillation solution to the
solar neutrino problem with large electron neutrino mixing. 
We remark that $U_{e2} = -0.016$ 
satisfies the CHOOZ electron neutrino survival probability 
bound ($U_{e2}$ is the relevant mixing matrix element, since 
$\Delta m_{12}^2 \sim \Delta m_{23}^2 \gg \Delta m_{13}^2$).

It is also possible to obtain a small mixing angle SMA-MSW 
solution to the solar neutrino problem, with a different 
choice of charges for $\Delta$:
\begin{equation}
\vec{Q}_{\Delta} = (-\frac{1}{2}, -\frac{2}{3}, -\frac{1}{6},0)
\end{equation} 
which gives the quasi-diagonal neutrino mass matrix
\begin{equation}
M_{\nu} \sim \langle\Delta^0\rangle\left ( \begin{array}{ccc}
W^4 T \xi^2 \psi^2 & \: W T^2 \xi^3  &  \: T^3 \xi^2 \psi \\
W T^2 \xi^3  & W T^2  & W T \xi^2 \\
T^3 \xi^2 \psi & W T \xi^2  &  \xi \psi \\
\end{array}
\right).
\end{equation}
The mixing matrix $U_{\nu}$ for this mass matrix is given by,
\begin{equation}
U_{\nu} \sim \left ( \begin{array}{ccc}
1 & \xi^3 & - T^3 \xi \\
-\xi^3 & 1 & \frac{WT\xi}{\psi}\\
-T^3 \xi & - \frac{WT\xi}{\psi} & 1 \\
\end{array} \right).
\end{equation}
Thus  we obtain the lepton mixing matrix:
\begin{equation}
U = U_E^{\dagger} U_{\nu}  \sim \left( \begin{array}{ccc}
1 & -W\psi & \frac{\xi \psi^2}{T} \\
 \frac{\xi \psi^2 X}{T} & \frac{WT}{\xi\psi X} & -\frac{1}{X} \\
-\frac{\xi^3}{X}  & \frac{1}{X}  & \frac{WT}{\xi\psi X} \end{array}\right)
\sim \left( \begin{array}{ccc}
1 & 0.016 & 0.014\\
0.021 & 0.75 & -0.66\\
6 \times 10^{-4} & 0.66 & 0.75 \end{array}\right).
\end{equation}
Taking $\langle\Delta^0\rangle \sim 3$ eV, we then obtain 
suitable masses and mixings for the solution
of both the solar and atmospheric neutrino problems:
\begin{equation}
\sin^2 2\theta_{atm} \sim 1  \ \Delta m_{23}^2 \sim 
1 \times 10^{-3} \mbox{eV}^2 \quad
\sin^2 2\theta_{sun} \sim 1  \ \Delta m_{12}^2 \sim 
6 \times 10^{-6} \mbox{eV}^2
\end{equation}

We did not manage to find an LMA-MSW solution, which is 
favoured by the latest solar neutrino data from Sudbury 
and SuperKamiokande, using this approach. However, during this 
workshop, Holger, Yasutaka and I constructed a promising   
LMA-MSW solution \cite{fnt} using the extended version of the AGUT 
model with right-handed neutrinos and the usual see-saw 
mechanism. 

\section*{Acknowledgements}

I should like to thank PPARC for a travel grant and my collaborators, 
Jon Chkareuli, Holger Bech Nielsen and Yasutaka Takanishi, for many 
discussions.

\end{document}